\documentclass[pre,twocolumn,showpacs]{revtex4}
\usepackage{amsfonts,amssymb,amsmath,bm}
\usepackage{times}

\begin{document}

\title{Velocity statistics inside coherent vortices generated by the inverse cascade of $2d$ turbulence}

\author{I.V.Kolokolov and V.V.Lebedev}

\affiliation{Landau Institute for Theoretical Physics, RAS, \\
142432, Ak. Semenova 1-A, \\ Chernogolovka, Moscow region, Russia; \\
NRU Higher School of Economics, \\
101000, Myasnitskaya 20, Moscow, Russia. 
}

\begin{abstract}

We analyze velocity fluctuations inside  coherent vortices generated as a result of the inverse cascade in the two-dimensional ($2d$) turbulence in a finite box. As we demonstrated in \cite{16KL}, the universal velocity profile, established in \cite{14LBFKL}, corresponds to the passive regime of flow fluctuations. The property enables one to calculate correlation functions of the velocity fluctuations in the universal region. We present results of the calculations that demonstrate a non-trivial scaling of the structure function. In addition the calculations reveal strong anisotropy of the structure function.

\end{abstract}


\maketitle

\section{Introduction}

Effects of the counteraction of (relatively fast) turbulence fluctuations with a coherent (relatively slow) flow is one of the central problems of turbulence theory  \cite{townsend80}. Usually the fluid energy is transferred from the slow large-scale flow to turbulent pulsations \cite{Frisch}. However, in some cases the energy can go from small-scale fluctuations to the large-scale ones that can lead to formation of a non-trivial mean flow \cite{BE12}. Even basic problems such as to determine at which mean velocity turbulent fluctuations are sustained is still object of intense investigations \cite{avila2011}. There is still no consistent theory for the mean (coherent) profile coexisting with turbulent fluctuations, so that even the celebrated logarithmic law for the turbulent boundary layer is a subject of controversy \cite{deb}. Here, we consider an important case: two-dimensional ($2d$) turbulence in a restricted box where large-scale coherent structures are generated from small-scale fluctuations excited by pumping. This process occurs because in two dimensions the non-linear hydrodynamic interaction favors the energy transfer to larger scales \cite{67Kra,68Lei,69Bat}.

Already, the first experiments on $2d$ turbulence \cite{Som} have shown that in a finite box with small bottom friction, the energy transfer to large scales leads to the formation of coherent vortices. First numerical simulations \cite{Smith1,Smith2,Borue} also show appearing coherent vortices in $2d$ turbulence. Subsequent more pronounced numerical simulations \cite{Colm} and experiments \cite{Shats} demonstrated that these vortices have well-defined and reproducible mean velocity (vorticity) profiles. This profile is quite isotropic with a power-law radial decay of vorticity inside the coherent vortex. The profile in that region depends neither on the boundary conditions (no-slip in experiments, periodic in numerics) nor on the type of forcing (random in numerics versus parametric excitation or electromagnetic forcing in experiments). The same flow profile is formed both in the statistically stationary case where the mean flow level is stabilized by the bottom friction and in the case where the average flow is still not stabilized and increases as time runs.

In the paper \cite{14LBFKL} results of intensive simulations of $2d$ turbulence were reported, they demonstrated that the vortex polar velocity profile is flat in some interval of distances from the vortex center, that we call the universal interval. The mean vorticity in the interval is inversely proportional to the distance $r$ from the vortex center. In the same paper a theoretical scheme based on conservation laws and on symmetry arguments was proposed that explains the flat velocity profile. The scheme predicts the value of the polar velocity $U=\sqrt{3\epsilon/\alpha}$ (where $\epsilon$ is the energy production rate and $\alpha$ is the bottom friction coefficient), that is in excellent agreement with the numerics \cite{14LBFKL}.

In the work \cite{16KL} we performed an analytical investigation of the coherent vortex in the universal interval. As a result, we established that the flat velocity profile corresponds to the passive regime of  the flow fluctuations where their self-interaction can be neglected. The passive regime admits consistent analytical calculations that confirm validity of the value $U=\sqrt{3\epsilon/\alpha}$ for the polar velocity. Besides, we found expressions for the viscous core radius of the vortex and for the border of the universal region where the flat velocity profile is realized. The results reported in the work \cite{16KL} explain why no flat velocity profile was observed in early simulations \cite{Smith1,Smith2,Borue} and imply that at some conditions a large number of coherent vortices could appear instead of a few vortices in numerics \cite{14LBFKL,Colm} and experiment \cite{Shats}.

In this paper we examine the spatial structure of the flow fluctuations. The passive nature of the fluctuations admits a detailed analytical analysis. We find the pair correlation functions of the velocity fluctuations in the universal interval at scales less than the distance $r$ from the vortex center and larger than the pumping length. There the correlation function possesses a definite scaling, the scaling is strongly anisotropic. The structure function of the velocity in the range is a linear function of the separation between the points. If the dissipation is strong enough, then it can restrict this region of the linear profile from above. At the end of the paper we discuss applicability conditions of the results and possible extensions of our scheme.

\section{General Relations}
\label{sec:general}

We consider the case where $2d$ turbulence is excited in a finite box of size $L$ by an external forcing. It is assumed to be a random quantity with homogeneous in time and space statistical properties. We assume also that correlation functions of the pumping force are isotropic. The main object of our investigation is the stationary (in the statistical sense) turbulent state caused by such forcing. To excite turbulence the forcing should be stronger than dissipation related both to the bottom friction and the viscosity at the pumping scale. That implies that the characteristic velocity gradient of the fluctuations produced by the forcing should be much larger than the flow damping at the pumping scale. The velocity gradient is estimated as $\epsilon^{1/3}k_f^{2/3}$, where $\epsilon$ is the energy flux (energy production rate per unit mass) and $k_f$ is the characteristic wave vector of the pumping force. Thus we arrive at the inequalities
 \begin{equation}
 \epsilon^{1/3}k_f^{2/3} \gg \alpha, \nu k_f^2.
 \label{weakness}
 \end{equation}
Here $\alpha$ is the bottom friction coefficient and $\nu$ is the kinematic viscosity coefficient, therefore $\nu k_f^2$ is the viscous damping rate at the pumping scale $k_f^{-1}$. In simulations, hyperviscosity is often used. In the case the inequalities (\ref{weakness}) are still obligatory for exciting turbulence, where $\nu  k_f^2$ has to be substituted by the hyperviscous damping rate at the pumping scale $k_f^{-1}$.

If the inequalities (\ref{weakness}) are satisfied then turbulence is excited in the box and random pulsations of different scales are formed due to non-linear hydrodynamic interaction. The pumped energy  flows to larger scales whereas the pumped enstrophy flows to smaller scales \cite{67Kra,68Lei,69Bat}. Thus two cascades are formed: the energy cascade (inverse cascade) realized at scales larger than the forcing scale $k_f^{-1}$ and the enstrophy cascade realized at scales smaller than the forcing scale $k_f^{-1}$. In an unbound $2d$ system the inverse energy cascade is terminated by the bottom friction at the scale
 \begin{equation}
 L_\alpha = \epsilon^{1/2}\alpha^{-3/2},
 \label{Lalpha}
 \end{equation}
where a balance between the energy flux $\epsilon$ and the bottom friction is achieved. The enstrophy cascade is terminated by viscosity (or hyperviscosity) \cite{BE12}.

In a finite box the above two-cascade picture is realized if the box size $L$ is larger than $L_\alpha$. Here we consider the opposite case $L<L_\alpha$. Then the energy, transferred by the nonlinearity to the box size $L$ by the inverse cascade, is accumulated there giving rise to a mean (coherent) flow. We analyze the statistically stationary case where the mean flow is already formed and stabilized by the bottom friction. To describe the flow, we use the Reynolds decomposition, that is the flow velocity is presented as the sum $\bm V+\bm v$ where $\bm V$ is the velocity of the coherent flow and $\bm v$ represents velocity fluctuations on the background of the coherent flow. Let us stress that $\bm V$ is an average over time, it possesses a complicated spatial structure.

As numerics and experiment show, the coherent flow contains some vortices separated by a hyperbolic flow. The characteristic velocity $V$ of the coherent motion can be estimated as $V\sim \sqrt{\epsilon/\alpha}$. The estimate is a consequence of the energy balance: in the stationary case the incoming energy rate $\epsilon$ is equal to the bottom friction rate. The characteristic mean vorticity in the hyperbolic region is estimated as $\Omega\sim L^{-1}\sqrt{\epsilon/\alpha}$. However, inside the coherent vortices the mean vorticity $\Omega$ is much higher than the estimate \cite{14LBFKL,Colm,Shats}. The maximal value of the mean vorticity $\Omega$ is achieved in the viscous core of the vortex. The radius of the core can be estimated as $(\nu/\alpha)^{1/2}$ \cite{16KL}.

\section{Coherent Vortex}
\label{sec:vortex}

Here we examine the flow inside the coherent vortex. We attach the origin of our reference system to the vortex center that is determined as the point of maximum vorticity. The definition corresponds to the procedures used in the works \cite{14LBFKL,Colm,Shats} to establish the mean vortex profile. The position of the vortex center fluctuates, in the laboratory experiments it fluctuates near a fixed position determined by the cell geometry. For the periodic setup (used in the numerics) the vortex center can shift essentially from its initial position, and only the average relative position of the vortices is fixed. The reference system is not inertial, and the velocity of the vortex center is subtracted from the flow velocity in the system. However, the flow vorticity in the reference system coincides with one in the laboratory reference system.

As it was established experimentally and numerically \cite{Colm,Shats,14LBFKL}, in the chosen reference system the mean flow possesses the axial symmetry. Such flow can be characterized by the polar velocity $U$ dependent on the distance $r$ from the vortex center. Then the mean vorticity is calculated as $\Omega=\partial_r U +U/r$. To obtain an equation for the profile $U(r)$, one has to use the complete Navier-Stokes equation. Assuming that the average pumping force is zero one finds after averaging the Reynolds equation \cite{MoYa}. Outside the viscous core where the viscous term is irrelevant we arrive at
 \begin{equation}
 \alpha U=-\left(\partial_r+\frac{2}{r}\right)\langle uv \rangle,
 \label{Ueq1}
 \end{equation}
where $v$ and $u$ are radial and polar components of the velocity fluctuations and angular brackets mean time averaging.

To analyze the flow fluctuations inside the vortex, it is convenient to start from the equation for the fluctuating vorticity $\varpi$
 \begin{equation}
 \partial_t\varpi+(U/r)\partial_\varphi\varpi+v\partial_r\Omega
 +\nabla\left(\bm v \varpi-\langle \bm v \varpi \rangle\right)
 =\phi-\hat\Gamma \varpi,
 \label{vact}
 \end{equation}
that is obtained from the same Navier-Stokes equation. Here $\varphi$ is polar angle, $\phi$ is curl of the pumping force, $\bm v$ is fluctuating velocity, and the operator $\hat\Gamma$ presents dissipation including some terms. Among the terms are  the bottom friction $\alpha$ and the viscosity term, $ - \nu \nabla^2$. For the case of hyperviscosity the last contribution to $\hat\Gamma$ is substituted by $(-1)^{p+1} \nu_p (\nabla^2)^p$ where $p$ is an integer number. An additional contribution to $\hat\Gamma$ is related to the non-linear interaction of the fluctuations. Though the interaction is weak, it could be larger than $\alpha$ and $-\nu\nabla^2$ because of the smallness of the contributions.

After solving the equation (\ref{vact}) one can restore the fluctuating velocity from the relation $\varpi=\partial_r v +v/r-\partial_\varphi u/r$ and the incompressibility condition $\partial_r u +u/r +\partial_\varphi v/r=0$. The scheme enables one to avoid calculation of the pressure, that is related to the velocity by a non-local relation. However, at restoring the velocity from the vorticity we still encounter non-local expressions. 

\section{Universal interval}

Further we consider the region outside the vortex core where the coherent velocity gradient is large enough,
 \begin{equation}
 U/r \gg \epsilon^{1/3}k_f^{2/3}.
 \label{passiver}
 \end{equation}
In this case fluctuations in the interval of scales between the pumping scale $k_f^{-1}$ and the radius $r$ are strongly suppressed by the coherent flow. The inequality (\ref{passiver}) means that the mean velocity gradient $U/r$ is larger than the gradient of the velocity fluctuations in the region at all scales larger than $k_f^{-1}$. Therefore the passive regime is realized there, that is the self-interaction of the velocity fluctuations is weak. The interval of scales outside the vortex core where the inequality (\ref{passiver}) is satisfied will be called further the universal interval of scales..

Moreover, the passive regime is realized for scales smaller than the pumping scale $k_f^{-1}$. Indeed, in the direct cascade the velocity gradients can be estimated as $\epsilon^{1/3}k_f^{2/3}$, upto logarithmic factors weakly dependent on scale, see \cite{71Kra,94FL,11FL}. Therefore the inequality (\ref{passiver}) means domination of the coherent velocity gradient in the interval of scales where the direct cascade would realize. The passive regime can be consistently analyzed. Then one neglects the nonlinear in the velocity fluctuations term in Eq. (\ref{vact}) staying with a linear equation for the vorticity fluctuation $\varpi$. The equation enables one to express $\varpi$ in terms of the pumping $\phi$ and then to calculate correlation functions of $\varpi$ via the correlation functions of $\phi$.

Further we focus on the case where the pumping $\phi$ is short correlated in time and has Gaussian statistics. Direct calculations \cite{16KL} show that in this case
 \begin{equation}
 \langle uv \rangle  =\epsilon/\Sigma,
 \label{uvuni}
 \end{equation}
where $\Sigma$ is the local shear rate of the coherent flow
 \begin{equation}
 \Sigma=r\partial_r\left(U/r\right)
 =\partial_r U -U/r.
 \label{shear}
 \end{equation}
The expression (\ref{uvuni}) is derived at the condition $\Sigma\gg \Gamma_f$, where $\Gamma_f$ is the damping of the velocity fluctuations at the pumping scale. Validity of the condition is guaranteed by the inequalities (\ref{weakness},\ref{passiver}). Some additional condition $\nu k_f^2\gg\alpha$ is needed for validity of the expression (\ref{uvuni}), the inequality is assumed to be satisfied in our scheme. (Note that the inequality is satisfied in numerics \cite{14LBFKL}.) The opposite case needs some additional analysis that is out of scope of our work.

Substituting the expression (\ref{uvuni}) into Eq. (\ref{Ueq1}) one finds a solution
 \begin{equation}
 U=\sqrt{3\epsilon/\alpha}, \qquad
 \Sigma=-U/r,
 \label{sqrt}
 \end{equation}
for the mean profile. Thus we arrive at the flat profile of the polar velocity found in Ref. \cite{14LBFKL} and confirmed analytically in Ref. \cite{16KL}. It is characteristic of the universal region.

The left-hand side of the inequality (\ref{passiver}) diminishes as $r$ grows. Therefore it is broken at some $r\sim R_u$. Substituting the expression (\ref{sqrt}) into Eq. (\ref{passiver}) one obtains
 \begin{equation}
 R_u = L_\alpha^{1/3}k_f^{-2/3}
 =\epsilon^{1/6}\alpha^{-1/2}k_f^{-2/3}.
 \label{scen2}
 \end{equation}
Note that $R_u$ can be larger or smaller than the box size $L$, depending on the system parameters. The case $R_u>L$ is, probably, characteristic of the numerics \cite{Colm} and the experiments \cite{Shats}, then the passive regime is realized everywhere in the box. In contrast, in numerics \cite{14LBFKL} the universal region is relatively small, $R_u<L$, and is well separated from the outer region, that is not completely passive.

\section{Vorticity fluctuations}

Since the flow fluctuations inside the universal region are passive we can use the linearized version of the equation (\ref{vact})
 \begin{equation}
 \partial_t\varpi+(U/r)\partial_\varphi\varpi+v\partial_r\Omega
 +\hat\Gamma \varpi =\phi,
 \label{vpas}
 \end{equation}
Since the pumping is assumed to be short correlated in time, its statistics is determined by the pair correlation function
 \begin{equation}
 \langle \phi(t,\bm k) \phi(t',\bm k')\rangle
 =2(2\pi)^2\epsilon\delta(\bm{k}+\bm{k}')\delta(t-t')k^2\chi(\bm{k}),
 \label{pum}
 \end{equation}
for the space Fourier transform of $\phi$. The function $\chi(\bm{k})$ has a profile with the characteristic pumping wave vector $k_f$ and is normalized:
 \begin{equation}
 \int\frac{d^2\bm{k}}{(2\pi)^2}\chi(\bm{k})=1.
 \label{normachi}
 \end{equation}
Then $\epsilon$ is the energy (per unit mass, per unit time) pumped to the system, that is the energy flux.

We analyze the fluctuations near a radius $r=r_0$ with scales much smaller than the radius. Then the shear approximation for the mean velocity can be used. We pass to the reference system rotating with the angular velocity $\Omega(r_0)$ and expand all terms in the equation (\ref{vpas}) in $x_1=r-r_0$ and $x_2=r_0\varphi$. We assume that the parameter $(k_f r)^{-1}$ is small. Then the term $v\partial_r\Omega$ in Eq. (\ref{vpas}) can be discarded and we end up with the following equation:
 \begin{equation}
 \partial_t\varpi+\Sigma x_1\partial_2\varpi+\hat{\Gamma}\varpi=\phi.
 \label{vispas}
 \end{equation}
Let us rewrite the evolution equation (\ref{vispas}) for the spatial Fourier components of the vorticity $\varpi_{\bm{k}}$:
 \begin{equation}
 \partial_t\varpi(\bm{k})-\Sigma k_2\partial_{k_1}\varpi(\bm{k})
 +\Gamma(k)\varpi(\bm{k})=\phi(t,\bm k).
 \label{pass1}
 \end{equation}
Solving the evolution equation (\ref{pass1}) one obtains a formal solution
 \begin{eqnarray}
 \varpi(t,\bm k)=\int^t d\tau\, \phi\left[\tau,k_1+\Sigma(t-\tau)k_2,k_2\right]
 \label{passolv} \\
 \times\exp\left\{-\int\limits_\tau^t\,d\tau'\Gamma\left[\sqrt{\left(k_1+\Sigma (t-\tau')k_2\right)^2+k_2^2}\right]\right\},
 \nonumber
 \end{eqnarray}

Now we can find the simultaneous pair vorticity correlation function for the Fourier transform from Eq. (\ref{pum})
 \begin{eqnarray}
 \langle\varpi(t,\bm{k})\varpi(t,\bm{k}')\rangle
 =2(2\pi)^2\epsilon \delta(\bm{k}+\bm{k}')
 \nonumber \\
 \int\limits_0^\infty d\tau\, q^2 \chi(\bm q)
 \exp\left[-2\int_0^\tau d\tau'\,\Gamma(\bm q')\right].
 \label{corr1}
 \end{eqnarray}
Here
 \begin{equation}
 \bm q=(k_1+\Sigma\tau k_2,k_2),
 \label{defq}
 \end{equation}
and $\bm q'$ differs from $\bm q$ by a substitution $\tau\to\tau'$. The factor $\delta(\bm{k}+\bm{k}')$ in the expression (\ref{corr1}) reflects space homogeneity that is not destroyed by a shear flow.

Let us consider scales larger than $k_f^{-1}$, that is wave vectors $k\ll k_f$. At the condition the main contribution to the integral (\ref{corr1}) is gained from times $\tau\sim k_f/(\Sigma k_2)$. In the case $|k_2|\gg \Gamma_f k_f/ \Sigma$ the dissipation is irrelevant and the last exponential factor in Eq. (\ref{corr1}) can be substituted by unity. Here, as above, $\Gamma_f$ is the flow damping at the pumping scale. Passing then to the integration over the wave vector (\ref{defq}), one obtains
 \begin{eqnarray}
 \langle\varpi(t,\bm{k})\varpi(t,\bm{k}')\rangle
 = \delta(\bm{k}+\bm{k}')
 \frac{2(2\pi)^2\epsilon q_f}{\Sigma |k_2|}
 \label{vortcor} \\
 q_f=\int\limits_0^\infty dq_1\, q^2\chi(q).
 \label{defq2}
 \end{eqnarray}
Here, we replaced the lower integration limit $|k_1|$ in the integral (\ref{defq2}) by zero, since the integral is gained at $q\sim k_f\gg k_1$. The wave vector $q_f$ is of the order of the inverse pumping length.

There can exist an interval of the wave vectors $r^{-1}<|k_2|<\Gamma_f k_f/ \Sigma$ where the dissipation is relevant. Then the exponential factor in Eq. (\ref{corr1}) is relevant. Therefore the expression (\ref{vortcor}) should be corrected by an additional small factor $\exp(-A)$, $A\sim \Gamma_f k_f/(\Sigma |k_2|)$. Thus the vorticity correlations are strongly suppressed in the region of wave vectors.

\section{Velocity correlation functions}

Knowing the vorticity correlation function, one can calculate the velocity correlation functions using the relation
 \begin{equation}
 v_\alpha({\bm{k}})=i\epsilon_{\alpha\beta}\frac{k_\beta}{k^2}\varpi(\bm{k}),
 \label{vortvel}
 \end{equation}
valid for the Fourier transforms. If $k_f\gg|k_2|\gg \Gamma_f k_f/ \Sigma$ and $k_f\gg|k_1|$, then one finds from Eqs.  (\ref{vortcor},\ref{vortvel})
 \begin{eqnarray}
 \langle v({\bm{k}})v({\bm{k}'})\rangle=
 2(2\pi)^3 \delta(\bm{k}+\bm{k}')
 \frac{q_f\epsilon}{\Sigma}\frac{|k_2|}{{\bm{k}}^4 },
 \label{strf11} \\
 \langle u({\bm{k}})u({\bm{k}'})\rangle=
 2(2\pi)^3 \delta(\bm{k}+\bm{k}')
 \frac{q_f\epsilon}{\Sigma}\frac{k_1^2}{{\bm{k}}^4 |k_2|},
 \label{strf12} \\
 \langle v({\bm{k}})u({\bm{k}'})\rangle=
 -2(2\pi)^3 \delta(\bm{k}+\bm{k}')
 \frac{q_f\epsilon}{\Sigma}\frac{k_1 k_2}{{\bm{k}}^4 |k_2|}.
 \label{strf22}
\end{eqnarray}
If $r^{-1}<|k_2|<\Gamma_f k_f/ \Sigma$ then the expressions are strongly suppressed due to dissipation.

It follows from the expressions (\ref{strf11},\ref{strf22}) that the averages $\langle v^2 \rangle$ and $\langle u^2 \rangle$ are determined by the infrared integrals. Therefore
 \begin{eqnarray}
 \langle v^2 \rangle,\langle u^2 \rangle
 \sim \frac{k_f \epsilon}{\Sigma}r \quad \mathrm{if}\
 \Gamma_f k_f r \ll \Sigma,
 \label{aver1} \\
 \langle v^2 \rangle,\langle u^2 \rangle
 \sim \frac{\epsilon}{\Gamma_f} \quad \mathrm{if}\
 \Gamma_f k_f r \gg \Sigma.
 \label{aver2}
 \end{eqnarray}
The average $\langle uv \rangle$ needs an additional analysis \cite{16KL}. It shows, that the quantity is gained at small scales and is determined by the expression (\ref{uvuni}). 

A special problem is calculation of $\langle u_0^2 \rangle$ where $u_0$ is zero angular harmonics of the fluctuating polar velocity. It is accounted by absence in the equation for $u_0$ an advection term related to the average flow. Therefore the quantity $\langle u_0^2 \rangle$ is determined solely by the damping. Strictly speaking, calculation of $\langle u_0^2 \rangle$ is outside of our shear approximation. However, our logic can be easily extended to the case to obtain
 \begin{equation}
 \langle u_0^2 \rangle \sim 
 \frac{\epsilon}{k_f r \Gamma_f} 
 \label{aver3}
 \end{equation}
An explanation of the expression is based on the expression
 \begin{equation*}
 \langle u_0^2 \rangle  
 =\int \frac{d\varphi}{2\pi}
 \langle u(\bm r_1) u(\bm r_2)\rangle,
 \end{equation*}
where the points $\bm r_1$ and $\bm r_2$ are separated by the same distance $r$ from the vortex center and $\varphi$ is the angle between the vectors $\bm r_1$ and $\bm r_2$. The factor $\epsilon/\Gamma_f$ is the contribution to $\langle u_0^2 \rangle$ caused by the pumping that is effective if $\varphi \lesssim (k_f r)^{-1}$. The contribution (\ref{aver3}) should be taken into account besides (\ref{aver1}), the latter is related to the sum of non-zero angular harmonics. In the case (\ref{aver2}) the contribution (\ref{aver3}) is small in comparison with one related to non-zero harmonics.

The average $\langle u_0^2 \rangle$ was calculated previously in the paper \cite{16Falk} where the contribution related to the pumping was ignored and the non-linear effects were taken into account instead. The approach is correct outside the universal region, at $r>R_u$ where $R_u$ is determined by the expression (\ref{scen2}). At the border, where $r\sim R_u$, our estimate (\ref{aver3}) coincides with one obtained in  Ref. \cite{16Falk}.

It is worth to characterize scales where the expressions (\ref{strf11}-\ref{strf22}) are correct by the velocity structure functions. One finds
\begin{eqnarray}
S_{11}(x_1,x_2)=\langle[v(x_1,x_2)-v(0,0)]^2\rangle
\nonumber \\
=\frac{2q_f\epsilon}{\Sigma\pi}\int d^2\bm{k}\,\frac{|k_2|}{\bm{k}^4 } \left(1-e^{ik_1x_1+ik_2x_2}\right),
\label{str1}
\end{eqnarray}
correct if $k_f^{-1}\ll|x_1,x_2|\ll r, k_f^{-1}\Sigma/\Gamma_f$. Infrared divergence in the integral (\ref{str1}) can be regularized by substituting $k_2^2\to k_2^2+\mu^2$, where $\mu\sim 1/r$. A result of the integration can be expressed via the function
\begin{equation}
\mathcal{J}(z)=
\int\limits_0^\infty dq\,\frac{e^{-z}}{q^2 +\mu^2}
\approx\frac{\pi}{2\mu}+z\left[\Gamma_f -1+\ln(\mu z)\right].
\label{insi}
\end{equation}
Particularly, one finds
\begin{equation}
S_{11}=\frac{2q_f\epsilon}{\Sigma}\mathrm{Re}\left[ \frac{\pi}{2\mu}-
\mathcal{J}+x_1\partial_1 \mathcal{J}\right],
\label{str2}
\end{equation}
where $\mathcal{J}=\mathcal{J}(x_1-ix_2)$, Calculating the expression (\ref{str2}), one finds
\begin{equation}
S_{11}\approx\frac{2q_f\epsilon}{\Sigma}\left[ |x_1|+x_2\arctan \left(\frac{x_2}{|x_1|}\right)\right].
\label{strua1}
\end{equation}

Analogous expressions can be found for other components of the structure function:
 \begin{eqnarray}
 S_{22}=\left\langle[u(x_1,x_2)-u(0,0)]^2\right\rangle\approx
 \nonumber\\
 \frac{2q_f\epsilon}{\Sigma} \left[x_2 \arctan \left(\frac{x_2}{|x_1|}\right)-2|x_1|
 \ln\left(\mu\sqrt{x_1^2+x_2^2}\right)\right],
 \label{strua2}
 \end{eqnarray}
and
 \begin{eqnarray}
 S_{12}=\langle[v(x_1,x_2)-v(0,0)][u(x_1,x_2)-u(0,0)]\rangle
 \nonumber\\
 \approx-\frac{2q_f\epsilon}{\Sigma} x_1 \arctan \left(\frac{x_2}{|x_1|}\right).
 \label{strua3}
 \end{eqnarray}
In the region $|x_1|,|x_2|\gg k_f^{-1}\Sigma/\Gamma_f$ the pair correlation functions are strongly suppressed and the structure functions are dominating by the single-point averages.

\section{Discussion}

We analyzed correlations of the velocity fluctuations inside a coherent vortex generated as a result of the inverse cascade in a finite $2d$ cell. Our attention was concentrated on the universal region inside the vertex where the mean velocity has the flat profile. We analyzed the fluctuations on a distance $r$ from the vortex core and with scales less than $r$. The amplitude of the velocity fluctuations grows as the scale grow as in the traditional inverse cascade. However, the expressions (\ref{strua1},\ref{strua2},\ref{strua3}) demonstrate linear profile, that is different from the $2/3$ power law in the traditional inverse cascade. Let us stress also that in our case the fluctuations are strongly anisotropic. Note also that at some conditions viscous dissipation can come into game, that leads to suppressing the fluctuations at the largest scale (below $r$).

We performed our calculations in the reference system where origin is attached to the vortex center and rotating with the angular velocity $\Omega$ dependent on the radius $r$ and coinciding with angular velocity of the mean flow at the distance $r$. In this reference system the correlation time of the pumping attached to the bottom of the cell cannot be larger than $\Omega^{-1}$. That justifies our approach (where the pumping is assumed to be short correlated in time) since the angular velocity $\Omega$ is the largest characteristic rate in the universal region. Note also, that use of the rotating reference system implies an implicit angular averaging of the correlation functions (besides the time averaging).

The universal region is restricted from above by the radius (\ref{scen2}). At larger distances from the vortex center the flow fluctuations are not completely passive, and our scheme is, strictly speaking, incorrect. In this case the traditional inverse cascade is realized on scales smaller than $l$, where $l\sim \epsilon^{1/2} \Sigma^{-3/2}$ is determined by the balance between the effective shear rate $\Sigma$ of the mean flow and the characteristic velocity gradient in the inverse cascade. However, the flow fluctuations are passive at scales larger than $l$. That is the region where our scheme is applicable. And the only difference is that the role of the pumping length is played just by the scale $l$.

Probably, our results can be extended for some types of three-dimensional turbulent flows. Note, as an example, the turbulence excited at the fluid surface \cite{Kameke,Xia2} where the inverse cascade is observed. It is a subject of future investigations.

\acknowledgements

We thank valuable discussions with G. Boffetta and G. Falkovich. The work is supported by RScF grant 14-22-00259.

\end{document}